\documentclass[a4paper]{jpconf}
\usepackage{graphicx}
\usepackage{amsmath}
\usepackage{mathrsfs}
\usepackage{amsfonts}
\usepackage{dsfont}
\usepackage{feynmp}
\usepackage{caption}

\begin{document}

\title{Non-conventional mesons at PANDA}
\author{Francesco Giacosa}

\begin{abstract}
Non-conventional mesons, such as glueballs and tetraquarks, will be in the
focus of the PANDA experiment at the FAIR facility. In this lecture we
recall the basic properties of QCD and describe some features of
unconventional states.\ We focus on the search of the not-yet discovered
glueballs and the use of the extended Linear Sigma Model for this purpose,
and on the already discovered but not-yet understood $X,Y,Z$ states.
\end{abstract}

\address{Institute of Physics, Jan Kochanowski University, 25-406 Kielce, Poland
and  Institute for Theoretical Physics, J. W. Goethe University,
Max-von-Laue-Str. 1, 60438 Frankfurt am Main, Germany}

\section{Introduction}

Conventional mesons are bound states made by a quark and an antiquark. Yet,
since the very beginning of QCD the search for other possibilities has
attracted the attention of both experimentalists and theoreticians \cite%
{review}.

As a prominent example, glueballs were predicted long ago: they are (yet
hypothetical) bound states of solely gluons. Computer simulations of QCD on
the lattice have found a full spectrum of these states \cite%
{mainlattice,Morningstar}, but their firm experimental discovery has not yet
taken place. On the other hand, a plenty of mesons, denoted as $X,$ $Y,$ and 
$Z$ states, has been unambiguously discovered in the last decade in the
energy region of charm-anticharm and botton-antibotton states \cite%
{brambilla,braaten}.\ A clear explanation about the nature of these states
is still lacking (tetraquark and molecular hypotheses are a possibility).

The PANDA experiment at the FAIR facility \cite{panda} in Darmstadt/Germany
is designed to shed light on these questions. It is a proton-antiproton
scattering experiment in which the energy of the antiproton can be finely
tuned in such a way that a wide energy range in the charmonium region can be
scanned. Various resonances can be directly formed in proton-antiproton
fusion processes. Glueball, if existent, shall be found by this experiment.
On the other hand, many of the $X,$ $Y,$ and $Z$ states can also be formed
and/or produced with high statistics. In this lecture a theoretical view
concerning the search of non-conventional mesons at PANDA will be presented.

\section{Brief recall of the QCD Lagrangian and its symmetries}

The Lagrangian of Quantum Chromodynamics (QCD) reads 
\begin{equation}
\mathcal{L}_{QCD}=\sum_{i=1}^{N_{f}}\overline{q}_{i}(i\gamma ^{\mu }D_{\mu
}-m_{i})q_{i}-\frac{1}{4}G_{\mu \nu }^{a}G^{a,\mu \nu }\text{ , }D_{\mu
}=\partial _{\mu }-ig_{0}A_{\mu }  \label{lqcd}
\end{equation}%
where $G_{\mu \nu }^{a}=\partial _{\mu }A_{\nu }^{a}-\partial _{\nu }A_{\mu
}^{a}+g_{0}f^{abc}A_{\mu }^{b}A_{\nu }^{c},$ $A_{\mu }=A_{\mu }^{a}t^{a},$ $%
a,=1,...,N_{c}^{2}-1=8$, and $t^{a}$ and $f^{abc}$ are the generators and
the structure constants of the group $SU(N_{c}).$ $\mathcal{L}_{QCD}$
contains $N_{f}$ quark fields $q_{i}$ with corresponding bare masses $m_{i}$
($i=1,...,N_{f}$, $N_{f}$ is the number of quark flavors; in Nature $N_{f}=6$%
: u, d, s, c, b, t). Each quark flavor has three colors (red, green, blue).
A crucial property of Eq. (\ref{lqcd}) is that 3-gluon and 4-gluon vertices
are present: gluons --contrary to photons-- `shine in their own light'. We
list the main symmetries and their meaning.

(i) $\mathcal{L}_{QCD}$ is built under the requirement of invariance under 
\emph{local} transformations of the $SU(3)$ groups. This means that one can
rename at \emph{each space-time point} the color of a quark via a $SU(3)$
matrix $U_{C}\equiv U_{C}(x)$ ($U_{C}U_{C}^{\dagger }=1,$ $\det U_{C}=1$)
and of the gluon field as: 
\begin{equation}
q_{i}\rightarrow U_{C}q_{i}\text{ , }A_{\mu }\rightarrow U_{C}(A_{\mu
}-i\partial _{\mu }/g_{0})U_{C}^{\dagger }\text{ .}
\end{equation}%
One can, for instance, transform a blue quark in a red one in a certain
space-time region without changing the properties of the system.

(ii) When the bare quark masses are equal ($m_{1}=m_{2}=...=m_{N_{f}}$), $%
\mathcal{L}_{QCD}$ is invariant under the interchange of quark flavors via a 
$N_{f}\times N_{f}$ unitary matrix $U_{V}$ ($U_{V}U_{V}^{\dagger }=1$) as: $%
q_{i}\rightarrow U_{V,ij}q_{j}$. This is flavor symmetry, denoted also as $%
U(N_{f})_{V}$: in simple terms, it means that gluons are `democratic' and
couple to each quark flavor with the same strength. Thus, one can rename
also the quark flavor: it is allowed to interchange a $u$ quark with a $d$
quark, but this can be done only once for all space-time points (the
symmetry is global and not local). For $N_{f}=2$, $m_{u}\simeq m_{d}\simeq 5$
MeV: the symmetry $U(2)_{V}$ is well realized in Nature (this is isospin
symmetry, responsible for instance for the almost equal mass of the three
pions and of the proton and the neutron). For $N_{f}=3$ the bare strange
mass $m_{s}\simeq 100$ MeV is sizably larger than $m_{u}$ and $m_{d}$:
nevertheless, the emergence of flavor multiplets with strange mesons is
evident in the PDG, showing that an approximate $U(3)_{V}$ is also realized 
\cite{PDG}.

The limit in which all quark masses vanish ($m_{1}=...=m_{N_{f}}=0)$ is
called the \emph{chiral limit} and is important for the understanding of QCD
because additional symmetries are present: dilatation invariance and chiral
symmetry.

(iii) Dilatation symmetry and its anomalous breaking. In the chiral limit,
there is only one parameter in $\mathcal{L}_{QCD}$, the \emph{dimensionless}
coupling constant $g_{0}$. The theory is classically dilatation invariant.
However, upon quantization, a running coupling $g_{0}\rightarrow g(\mu )$
emerges. An ultraviolet cutoff $\Lambda _{UV}$ is introduced in the process
of regularization in such a way that $g_{0}$ is the coupling at this very
high energy scale: $g_{0}=g(\Lambda _{UV})$ . Then, a low-energy scale $%
\Lambda _{YM}$ emerges in the theory as%
\begin{equation}
g^{2}(\mu )=\frac{g_{0}^{2}}{1+2bg_{0}^{2}\log \frac{\mu }{\Lambda _{UV}}}%
\rightarrow \Lambda _{YM}=\Lambda _{UV}e^{-1/(2bg_{0}^{2})}\text{ with }b=%
\frac{33-2N_{f}}{48\pi ^{2}}\text{.}  \label{ta}
\end{equation}%
Numerically, $\Lambda _{YM}\simeq 250$ MeV: all quantities in QCD depend
crucially on it.

(iv) Chiral symmetry and its spontaneous breaking. One splits the quark
field into the right-handed and left-handed components: $%
q_{i}=q_{i,L}+q_{i,R}=P_{L}q_{i}+P_{R}q_{i}$ with $P_{R(L)}=\frac{1}{2}%
\left( 1\pm \gamma ^{5}\right) .$ $\mathcal{L}_{QCD}$ is \emph{separately }%
invariant under rotations of right-handed quarks and left-handed quarks 
\begin{equation}
q_{i}=q_{i,L}+q_{i,R}\rightarrow U_{L,ij}q_{j,L}+U_{R,ij}q_{j,R}\text{ ,}
\end{equation}%
where $U_{R}$ and $U_{L}$ are two independent unitary matrices. Such a
chiral transformation is also denoted as $U(N_{f})_{R}\times U(N_{f})_{L}$.
Chiral transformation reduces to a flavor one for $U_{V}=U_{L}=U_{R}.$
Conversely, the case $U_{A}=U_{L}=U_{R}^{\dagger }$ is called axial
transformation $U(N_{f})_{A}$ (which is not a group!), which mixes states
with different parity, as pseudoscalar and scalar mesons and vector with
axial-vector ones. This symmetry is not realized in the hadronic spectrum 
\cite{PDG} because it is spontaneously broken by the nonperturbative QCD
vacuum. As a consequence the quarks develop -even in the chiral limit- a
large constituent (or effective) mass $m\rightarrow m^{\ast }\simeq \Lambda
_{YM}$.

\section{Mesons}

Quarks and gluons are the basic degrees of freedom of the $QCD$ Lagrangian
of Eq. (\ref{lqcd}). However, these are \emph{not} the asymptotic states
that we measure in our detectors. Namely, quarks and gluons are \emph{%
confined} into hadrons, where each hadron is \emph{white} (i.e., invariant
under the local color transformation introduced in\ Sec. 2).

We use the following \textit{definition}: `A meson is a strongly interacting
particle (a hadron) with integer spin'. This definition is consistent with
the PDG \cite{PDG}, in which all mesons are listed together independently of
their inner structure.

\subsection{Quark-antiquark mesons}

A \textit{conventional meson} is a meson constructed out of a quark and an
antiquark. Although it represents only one of (actually infinitely many)
possibilities to build a meson, the vast majority of mesons of the PDG can
be correctly interpreted as belonging to a quark-antiquark multiple \cite%
{PDG} (see also the results of the quark model \cite{isgur}).

Mesons can be classified by their spatial angular momentum $L,$ the spin $S,$
the total angular momentum $J$ and by parity $P$ and charge conjugation $C$
(summarized in $J^{PC}$). The lightest mesons are pseudoscalar states with $%
L=S=0\rightarrow J^{PC}=0^{-+}.$ Indeed, the pions and the kaons are
pseudoscalar (quasi-)Goldstone bosons emerging upon the spontaneous breaking
of chiral symmetry. As an example, we write down the wave function for the
state $K^{+}$ (radial, angular, spin, flavor, color): 
\begin{equation}
\left\vert K^{+}\right\rangle =\left\vert n=1\right\rangle \left\vert
L=0\right\rangle \left\vert S=0(\uparrow \downarrow -\downarrow \uparrow
)\right\rangle \left\vert u\bar{s}\right\rangle \left\vert \bar{R}R+\bar{G}G+%
\bar{B}B\text{ }\right\rangle \text{.}
\end{equation}

For $L=0,$ $S=1$ one constructs the vector mesons (such as $\rho $ and $%
\omega $), for $L=S=1$ one has three multiplets: tensor mesons $%
J^{PC}=2^{++},$ axial-vector mesons $J^{PC}=1^{++}$ and scalar mesons $%
J^{PC}=0^{++}$ (scalar states are in the center of a long debate, see e.g.
Refs. \cite{close,lowscalars,tetraquark} and refs. therein). By further
increasing $L$ one can obtain many more multiplets \cite{PDG}.

It is interesting to notice that the quantum numbers $J^{PC}=0^{+-}$ \emph{%
cannot} be obtained in a quark-antiquark system, but is possible for
unconventional mesonic states (such as glueballs). The experimental
discovery of mesons with such\emph{\ exotic quantum numbers }naturally
points to a non-quarkonium inner structure.

\subsection{Glueballs search and the eLSM}

According to lattice QCD many glueballs should exist \cite%
{mainlattice,Morningstar}, but up to now no glueball state has been
unambiguously identified (although for some of them some candidates exist).

A suitable theoretical framework to study the decays of glueballs is the
so-called extended linear Sigma Model (eLSM), which is an effective model of
QCD built accordingly to the two fundamental symmetries mentioned in\ Sec.
2: chiral symmetry and dilatation invariance. The former is spontaneously
broken by a Mexican-hat potential, the latter explicitly broken in order to
mimic the trace anomaly of QCD of Eq. (\ref{ta}), see Ref. \cite{migdal}. As
a consequence, the eLSM Lagrangian contains only a finite number of terms.
Moreover, (axial-)vector d.o.f. are included from the very beginning. The
eLSM was first developed for $N_{f}=2$ in Refs. \cite{susden,staninf2}, for $%
N_{f}=3$ in\ Refs. \cite{dick,staninew}, and for $N_{f}=4$ in Ref. \cite{nf4}%
. In particular, in Ref. \cite{dick} a fit to many experimental data was
performed and a good description of low-hadron phenomenology (up to about $%
1.7$ GeV) was obtained. Here we briefly recall the main results concerning
glueballs.

\emph{The scalar glueball} is the lightest gluonic state predicted in\ QCD
and is naturally an element of the eLSM as the excitation of the dilaton
field\ \cite{staninf2,dick,staninew}. The result of the recent study of Ref. 
\cite{staninew} shows that the scalar glueball is predominantly contained in
the resonance $f_{0}(1710)$, in agreement with the lattice result of Ref. 
\cite{chen}. The eLSM makes predictions for the lightest (and peculiar)
glueball state in a chiral framework, completing previous phenomenological
works on the subject \cite{close}.

\emph{The pseudoscalar glueball} is related to the chiral anomaly and
couples in a chirally invariant way to light mesons \cite{psg}, where it was
shown that it decays predominantly in $\pi \pi K$ ($50\%$ of all decays into
(pseudo)scalar mesons) and that it does \emph{not} decay in $\pi \pi \pi $:
these are simple and testable theoretical predictions which can be helpful
in the experimental search at the PANDA experiment, where the pseudoscalar
glueball can be directly formed in proton-antiproton fusion process.

A similar program can be carried out for a tensor glueball with a mass of
about $2.2$ GeV, e.g. Ref. \cite{tensor}, as well as for heavier glueballs,
such as the (pseudo)vector ones.

\subsection{X,Y,Z states and other non-quarkonium candidates}

The discovery in the last years of a plenty of enigmatic resonances -the so
called $X,Y,Z$ states- shows that there are now many candidates of
resonances beyond the standard quark-antiquark picture, see e.g. \cite%
{brambilla,braaten} ($X(3872)$ was the first to be experimentally found by
BELLE in 2003).The interpretation of these states is subject to ongoing
debates: tetraquarks and molecular interpretations are at the top of the
list, but it is difficult to distinguish among them \cite{braaten,maianix}.
Moreover, distortions due to quantum fluctuations of nearby threshold(s)
take place and make the understanding of these resonances more complicated 
\cite{coitox}. Remarkably, the $Z$ states are \emph{charged} states in the
charmonium region: a system made of four quarks is here necessary to
understand them since a charmonium is necessarily chargeless (see e.g. Ref. 
\cite{maianiz}).

There are also other mesonic states which are not yet understood. An example
is the strange-charmed scalar state $D_{S0}(2317)$, which is too light to be
a $c\bar{s}$ state and could be a four-quark or a dynamically generated
state. Historically, the scalar mesons below 1 GeV were among the first to
be interpreted as non-quarkonium objects, but as a nonet of tetraquarks \cite%
{tetraquark} or as dynamically generated states \cite{lowscalars}.

\section{PANDA: formation and production of mesons}

In the future PANDA experiment at the FAIR facility in Darmstadt \cite{panda}%
, antiprotons reach a three-momentum $\mathbf{k}$ in the range $\left\vert 
\mathbf{k}\right\vert =2.2$-$10$ GeV and hit protons at rest. We consider
here the case in which the proton and the antiproton completely annihilate
and generate a particle $X$ (as for instance a glueball) with mass $m_{X}$.
The four-momentum of the antiproton reads $k_{\bar{p}}=(E_{\bar{p}}=\sqrt{%
\mathbf{k}^{2}+m_{p}^{2}},\mathbf{k})$ while that of the proton is $%
k_{p}=(m_{p},\mathbf{0}).$ By denoting $k_{X}=(\sqrt{\mathbf{k}^{2}+m_{X}^{2}%
},\mathbf{k})$ as the four-momentum of $X,$ we obtain out of $k_{p}+k_{\bar{p%
}}=k_{X}$ that: 
\begin{equation}
m_{X}=\sqrt{2m_{p}\left( m_{p}+E_{\bar{p}}\right) }=2.25\text{-}4.53\text{
GeV.}
\end{equation}%
By looking at the lattice spectrum of Ref. \cite{mainlattice}, we realize
that -besides the scalar glueball which is too light- all non-exotic
glueball states could be directly formed at the PANDA experiment. This
represents a clean environment to study experimentally their decays.
Glueballs with exotic quantum numbers (called oddballs) cannot be directly
formed because a proton-antiproton system undergoes the same limitations of
a quark-antiquark system for what concerns $J^{PC}$ quantum numbers.
Nevertheless, oddballs will also be produced together with other resonances
and will be studied as well. Besides the search for glueballs, all mesonic
states discussed above will be experimentally investigated.

\section{Conclusions and outlook}

In this work we have given a brief overview of some aspects of the
theoretical as well as experimental search for unconventional mesons. The
PANDA experiment will play a decisive role in the future of hadron physics,
since it will help to clarify many open questions of hadron spectroscopy in
general and of exotic mesons in particular. The search for and hopefully the
firm discovery of glueballs (and hybrids), as well as the confirmation and
measurement with high statistics of the $X,Y,Z$ states will be important
milestones toward a better understanding of QCD.

\section*{Acknowledgments}

The author thanks all the members of the group for chiral field theories of
Frankfurt for many valuable cooperations and discussions.

\end{document}